\documentclass[%
 reprint,
 amsmath,amssymb,
 aps,
 pre,
 longbibliography
]{revtex4-1}

\usepackage{CJK}
\usepackage{graphicx}% Include figure files
\usepackage{dcolumn}% Align table columns on decimal point
\usepackage{bm}% bold math
\usepackage{color}
\usepackage{amsmath}
\begin{document}

\begin{CJK*}{UTF8}{} % Use default fonts from CJK (see below)

%\preprint{APS/123-QED}

\title{Capillary Forces on a Small Particle at a Liquid-Vapor Interface: Theory and Simulation}
%\thanks{A footnote to the article title}%

\author{Yanfei Tang ({\CJKfamily{gbsn}唐雁飞})}
\affiliation{Department of Physics, Center for Soft Matter and Biological Physics,
and Macromolecules Innovation Institute, Virginia Polytechnic Institute and State University,
Blacksburg, Virginia 24061, USA}
\author{Shengfeng Cheng ({\CJKfamily{gbsn}程胜峰})}
\email{chengsf@vt.edu}
\affiliation{Department of Physics, Center for Soft Matter and Biological Physics,
and Macromolecules Innovation Institute, Virginia Polytechnic Institute and State University,
Blacksburg, Virginia 24061, USA}

\date{\today}% It is always \today, today,
             %  but any date may be explicitly specified

\begin{abstract}
We study the meniscus on the outside of a small spherical particle with radius $R$ at a liquid-vapor interface. The liquid is confined in a cylindrical container with a finite radius $L$ and has a contact angle $\pi/2$ at the container surface. The center of the particle is placed at various heights along the central axis of the container. By varying $L$, we are able to systematically study the crossover of the meniscus from nanometer to macroscopic scales. The meniscus rise or depression on the particle is found to grow as $\ln (2L/R)$ when $R\ll L\ll \kappa^{-1}$ with $\kappa^{-1}$ being the capillary length and saturate to a value predicted by the Derjaguin-James formula when $R \ll \kappa^{-1} \ll L$. The capillary force on the particle exhibits a linear dependence on the particle's displacement from its equilibrium position at the interface when the displacement is small. The associated spring constant is found to be $2\pi\gamma\ln^{-1} (2L/R)$ for $L\ll \kappa^{-1}$ and saturates to $2\pi\gamma\ln^{-1} (3.7\kappa^{-1}/R)$ for $L\gg \kappa^{-1}$. At nanometer scales, we perform molecular dynamics simulations of the described geometry and the results agree well with the predictions of the macroscopic theory of capillarity. At micrometer to macroscopic scales, comparison to experiments by Anachkov \textit{et al.} [Soft Matter {\bf 12}, 7632 (2016)] shows that the finite span of a liquid-vapor or liquid-liquid interface needs to be considered to interpret experimental data collected with $L \sim \kappa^{-1}$.
 
%\begin{description}
%\item[Usage]
%Secondary publications and information retrieval purposes.
%\item[PACS numbers]
%May be entered using the \verb+\pacs{#1}+ command.
%\item[Structure]
%You may use the \texttt{description} environment to structure your abstract;
%use the optional argument of the \verb+\item+ command to give the category of each item. 
%\end{description}
\end{abstract}

%\pacs{Valid PACS appear here}% PACS, the Physics and Astronomy
                             % Classification Scheme.
%\keywords{Suggested keywords}%Use showkeys class option if keyword
                              %display desired
\maketitle

\end{CJK*}

%\tableofcontents

\section{introduction} \label{sec:intro}

Recently, the drying of colloidal suspensions has attracted great attention as it provides a facile procedure to generate dry colloidal films and superstructures with controlled arrangements of particles \cite{Zhou2017AdvMater}. To understand the structural formation in colloidal suspensions induced by solvent evaporation, much effort has been made to model such systems using molecular dynamics (MD) simulations \cite{Fortini2016,Fortini2017,Howard2017,Reinhart2017,Tang2018Langmuir,Statt2018arXiv}. In these simulations, one key aspect is the representation of the solvent. In a few works, the solvent is modeled explicitly as Lennard-Jones liquids \cite{Tang2018Langmuir,Statt2018arXiv}. However, such simulations are extremely expensive and the parameter space that can be explored is rather limited \cite{Tang2018Langmuir}. In others, an implicit solvent model is adopted and a liquid-vapor interface is modeled as a confining potential for all the solutes in the solution \cite{Fortini2016,Fortini2017,Howard2017,Reinhart2017}. Usually, a harmonic potential is used with the potential minimum indicating a particle's equilibrium position relative to the interface \cite{Pieranski1980}. The evaporation process of the solvent is mimicked by moving the interface in a controlled manner. In this moving interface method of modeling the evaporation process of a solvent, a spring constant has to be assumed in the harmonic potential to capture the confining effect of the interface on the particles in the liquid solvent. Though a harmonic potential for a particle adsorbed at an interface is intuitively sensible and has been widely used \cite{Pieranski1980,Colosqui2013}, there lacks a systematic physical interpretation of the associated spring constant. A deeper understanding is thus needed on the effective potential experienced by a particle when it is displaced out of its equilibrium position at a liquid-vapor interface.

Understanding the behavior of a particle at a liquid-vapor interface (or more generally, a fluid-fluid interface) is also important in many fields such as interfacial self-assembly of particles \cite{Lin2003,Dong2010,Ershov2013,Davies2014AdvMater}, emulsion and foam stabilization \cite{Stratford2005,Frost2011,Luu2013}, fabrication of colloidal gels \cite{Sanz2009}, interfacial particle adsorption \cite{Colosqui2013,Hua2018}, flotation processing of minerals~\cite{Scheludko1976}, and granular materials \cite{Mitarai2006}. A comprehensive review of this topic can be found in Ref.~\cite{Bresme2007}. Because of its practical importance, the detachment of a particle from a planar liquid surface has been studied for a long time~\cite{Scheludko1975, Scheludko1976, Rapacchietta1977cylinders, Rapacchietta1977spheres, Huh1976, OBrien1996, Pitois2002, Chateau2003, Garbin2012, Anachkov2016, Dutka2017, Zanini2018}. The quasistatic removal of a sphere from a liquid surface has a strong connection with the meniscus on the outside of a cylinder in a liquid bath, which is governed by the Young-Laplace equation. Huh and Scriven numerically studied this problem for an unbound liquid surface \cite{Huh1969}. A formula for the meniscus height was proposed by James for this case \cite{James1974}, which was actually suggested earlier by Derjaguin \cite{Derjaguin1946} (see Note at the end of Ref.~\cite{James1974}). We call this result the Derjaguin-James formula, which is very accurate for small cylinders. A similar formula for the meniscus height on a sphere at the surface of a large liquid bath has been widely used in later research~\cite{Scheludko1976}. Pitois and Chateau studied the work of detachment of removing a small particle from an interface both experimentally and analytically using a theory based on the Derjaguin-James formula~\cite{Pitois2002, Chateau2003}. Anachkov \textit{et al.} recently refined Pitois and Chateau's theory by correcting the critical central angle at which a capillary bridge ruptures and compared the theory with experimental data collected with a colloidal-probe atomic force microscope (AFM) \cite{Anachkov2016}.

Pulling a small sphere from a liquid surface can be used as a technique known as sphere tensiometry to measure the surface tension of the liquid and its contact angle on the surface of the sphere \cite{Scheludko1975, Huh1976}. This method is based on the fact that surface tension is the physical origin of the capillary force on a particle that controls its detachment behavior. Depending on the size of the particle, gravity and buoyancy force may also come into play \cite{Rapacchietta1977cylinders, Rapacchietta1977spheres}. To the best of our knowledge, in most studies reported so far on particle detachment from a liquid surface, the surface was assumed to be unbound laterally \cite{Scheludko1975, Scheludko1976, Rapacchietta1977cylinders, Rapacchietta1977spheres, Huh1976, OBrien1996, Pitois2002, Chateau2003, Butt2016, Anachkov2016, Zanini2018}. This assumption is valid when the lateral size of the liquid bath, $L$, is much lager than the capillary length, $\kappa^{-1}$, of the interface. However, in recent AFM experiments the size of the particle is at the scale of micrometers and the lateral span of the meniscus can be comparable or even smaller than $\kappa^{-1}$ \cite{Anachkov2016}. De Baubigny \textit{et al.} showed that the lateral size of a liquid bath can affect the capillary force and meniscus rise on a nanofiber \cite{DeBaubigny2015}. Recently, we studied the meniscus on a small cylinder located at the center of a liquid-vapor interface that is confined in a cylindrical container of a finite radius $L$ \cite{Tang2018arXiv}. Our results show the crossover from nano-/micrometer scales where the meniscus rise grows with $L$ logarithmically to macroscopic scales where the meniscus rise saturates to a value predicted by the Derjaguin-James formula. Similar crossover is expected to occur for a small sphere as well.

In a seminal work on contact angle hysteresis, Joanny and de Gennes showed that the capillary force associated with contact line pinning on a defect exhibits a linear dependence on the deformation of the contact line, and the resulting spring constant has a logarithmic dependence on a length scale, which can be interpreted as the average distance between the defects \cite{Joanny1984}. An experiment by Nadkarni and Garoff on the contact line pinning on a single defect confirmed the theoretical prediction of Joanny and de Gennes and revealed a relation between the pinning of a contact line and the removal of a particle from a liquid surface \cite{Nadkarni1992}. A similar connection was discussed by O'Bien as well \cite{OBrien1996}. Later works by Preuss and Butt \cite{Preuss1998}, and by Ettelaie and Lishchuk \cite{Ettelaie2015, Lishchuk2016} showed that a linear force-displacement curve emerges not only for a particle detaching from a planar liquid surface but also from a surface with an overall curvature. A similar behavior was observed for spheroidal particles by Davies \textit{et al.} \cite{Davies2014}.

In this paper our goal is to study the force-displacement curve for a particle at a liquid-vapor interface with a finite lateral extent ranging from nanometer to macroscopic scales. To achieve this goal, we study a small particle with radius $R$ at a liquid-vapor interface with both the macroscopic theory of capillarity and MD simulations, the latter of which has been widely used recently to study capillary phenomena at nanometer scales \cite{Seveno2004,DeConinck2008, Seveno2013,Cheng2014,Cheng2016Langmuir}. In particular, we place the liquid in a cylindrical container with radius $L$ ($>R$) and the particle along the central axis of the container. The meniscus around and the capillary force on the particle are computed when it is placed at different height across the liquid-vapor interface. This geometry allows us to systematically explore the crossover from a region where $L \ll \kappa^{-1}$, and thus gravity can be ignored, to a region where $L \gtrsim \kappa^{-1}$ and gravity starts to play a role. We systematically study this crossover with the macroscopic theory of capillarity based on the Young-Laplace equation. At nanometer scales, we perform MD simulations and compare the simulation results on the meniscus profile and capillary force with the predictions of the macroscopic theory and find a good agreement between the two. At micrometer to macroscopic scales, we compare the theory with the experimental data by courtesy of Dr. Anachkov \cite{Anachkov2016} and show that the finite extent of the menisci involved in the experiments needs to be considered to understand the experimental results.

This paper is organized as follows. In Sec.~\ref{sec:theory} we present a complete theory of a meniscus on the outside of a sphere with the lateral span of the meniscus varying from nanometer to macroscopic scales. The MD simulation methods are introduced in Sec.~\ref{sec:simu}. In Sec.~\ref{sec:RD} we discuss and compare the results from the theory, simulations, and experiments. We conclude the paper in Sec.~\ref{sec:conclusions} by summarizing the results on the effective confining potential on a particle from a liquid-vapor interface, which provide a physical foundation of the moving interface method of modeling solvent evaporation.

\begin{figure*}[btp]
  \includegraphics[width = 0.8 \textwidth]{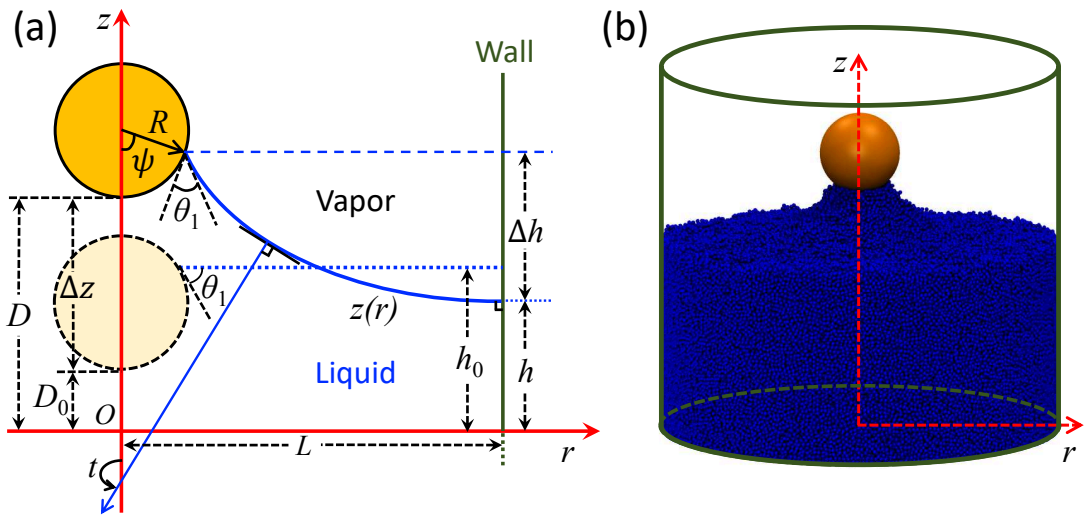}
  \caption{(a) Sketch of a particle at the center of the surface of a liquid bath in a cylindrical container: particle at an equilibrium height (dashed light yellow sphere) and being pulled upward (solid orange sphere). (b) Snapshot from MD simulations of a system with a particle pulled from a liquid-vapor interface.}
  \label{fg:main}
\end{figure*}

\section{Macroscopic theory of capillarity} \label{sec:theory}

\subsection{General Theory}

The geometry of the systems studied in this paper, as well as a snapshot from MD simulations, is sketched in Fig.~\ref{fg:main}, where a particle of radius $R$ straddles a liquid-vapor interface. The liquid bath is placed in a cylindrical container with radius $L$ and $z$-axis being its central axis, along which the particle's center is located. A bottom wall at $z=0$ is used to confine the liquid from below. To study the crossover from a system where $L$ is larger than but comparable to $R$ to a system where the meniscus is unbound (i.e., $L \rightarrow \infty$), we set the contact angle of the liquid on the container surface to be $\pi/2$. This choice guarantees that the liquid-vapor interface far away from the particle is flat when $L \rightarrow \infty$, which is expected for an unbound meniscus. The contact angle of the liquid on the particle surface is $\theta_1$. Here we do not consider any pinning effect of a contact line on the surface of either the particle or the container. When the particle is in its equilibrium location, the liquid-vapor interface is flat and intersects with the particle surface at a filling angle $\psi = \pi-\theta_1$, as shown in Fig.~\ref{fg:main}(a). We denote the height of this flat interface in equilibrium as $h_0$. The equilibrium position of the particle center is then $h_0 - R\cos \theta_1$ \cite{Bresme1998}.

When the particle is pulled upward or pressed downward vertically (i.e., along the central axis of the cylindrical container), the liquid-vapor interface will bend and form a meniscus, as shown in Fig.~\ref{fg:main} for pulling. If we denote the distance from the bottom of the particle to the bottom wall as $D$, then $D_0 = h_0 - R(1+\cos \theta_1)$ for a particle in its equilibrium location and $\Delta z \equiv D-D_0 = D - h_0 + R (1+\cos \theta_1)$ is the displacement of the particle from its equilibrium height. In this paper, our main goal is to understand how the capillary force on the particle depends on $\Delta z$. For this purpose, we must solve the meniscus profile, $z(r)$. For simplicity, we define $z_0(r) \equiv z(r) - h$, where $h$ is the height at which the meniscus meets the container surface. Then the range of $z_0$ is $[0,\Delta h]$ (or $[\Delta h,0]$), where $\Delta h$ is the amount of the meniscus rise (or depression).

Considering that a meniscus can develop a neck where $|\frac{\mathrm{d}z_0}{\mathrm{d}r}| = \infty$, the function $z_0(r)$ becomes double-valued for a range of $r$ near the neck. It is therefore more convenient to represent the meniscus as $r(z_0)$, which is always a single-valued function of $z_0\in [0,\Delta h]$ for a rising meniscus (or $z_0\in [\Delta h,0]$ for a depressed meniscus). The function $r(z_0)$ is the solution of a form of the Young-Laplace equation studied by Bashforth and Adams before \citep{Bashforth1883},
\begin{align}
\frac{r''}{(1 + r'^2)^{3/2}} - \frac{1}{r (1 + r'^2)^{1/2}}
= \frac{\Delta p}{\gamma} + \frac{\Delta \rho g z_0}{\gamma}~,
\label{eq:Young-Laplace}
\end{align}
where $r' \equiv \frac{\mathrm{d}r}{\mathrm{d}z_0}$, $r'' \equiv \frac{\mathrm{d}^2r}{\mathrm{d}z_0^2}$, $\Delta p$ is the pressure jump from the vapor to the liquid phase at $r=L$ (i.e., the pressure on the vapor side minus that on the liquid side across the liquid-vapor interface at $r=L$), $\gamma$ is the surface tension of the liquid, $\Delta \rho \equiv \rho_l - \rho_v$ is the difference of the liquid and vapor densities, and $g$ is the gravitational constant. Eq.~(\ref{eq:Young-Laplace}) is for a rising meniscus that may have a neck. To describe a depressed meniscus, the left hand side of Eq.~(\ref{eq:Young-Laplace}) should be multiplied with $-1$.

A physical solution of Eq.~(\ref{eq:Young-Laplace}) for a given $D$ needs to satisfy the constraint that the volume of the liquid bath,
\begin{align}
V &= \pi \int_{0}^{\Delta h} r^2(z_0) \, \mathrm{d}z_0   \nonumber \\
&+ \pi L^2 h - \frac{\pi R^3}{3}\left(2 - 3\cos \psi + \cos^3 \psi\right)~,
\label{eq:vol_bath}
\end{align}
is fixed at a constant set by parameters $h_0$, $L$, $R$, and $\theta_1$, which set up the physical problem at hand. Since for the particle in its equilibrium location, $\Delta h=0$, $h=h_0$, and $\psi = \pi - \theta_1$, the fixed volume is
\begin{align}
V = \pi L^2 h_0 - \frac{\pi R^3}{3}\left(2 + 3\cos \theta_1 - \cos^3 \theta_1\right)~.
\label{eq:vol_initial}
\end{align}
Eqs.~(\ref{eq:Young-Laplace})--(\ref{eq:vol_initial}) actually provide an implicit relation for the filling angle $\psi$, which in turn determines the meniscus profile on the outside of the particle. For a given $\Delta z$, we have
\begin{align}
D = \Delta z + h_0 - R (1+\cos \theta_1)~.
\label{eq:D}
\end{align}
The procedure of solving the meniscus profile for the given $\Delta z$ starts with an assumed filling angle $\psi$. Then Eq.~(\ref{eq:Young-Laplace}) is solved either analytically or numerically to obtain the meniscus profile, $r(z_0)$, including the meniscus height, $\Delta h$. The height of the liquid-vapor interface at $r=L$ is then given by
\begin{align}
h = D+R(1-\cos \psi) -\Delta h~.
\label{eq:interface_height_container}
\end{align}
With $r(z_0)$, $\Delta h$ and $h$ determined, the volume of the liquid bath can be computed with Eq.~(\ref{eq:vol_bath}) and compared to Eq.~(\ref{eq:vol_initial}) until for the given $\Delta z$ a filling angle $\psi$ is found to satisfy the volume constraint.

After the meniscus profile is determined at a given $\Delta z$, i.e., after the filling angle $\psi$ is found for the given $\Delta z$ using the self-consistent procedure described above, the total capillary force on the particle can be computed as
\begin{align} \label{eq:force}
F &= 2\pi\gamma R \sin \psi \sin(\theta_1 + \psi)   \nonumber \\
& - \left(\Delta p + \Delta \rho g \Delta h\right)  \pi R^2 \sin^2 \psi   \nonumber \\
&- \Delta \rho g \frac{\pi R^3}{3} \left(2 - 3\cos \psi + \cos^3 \psi\right)~,
\end{align}
where the first term is a direct contribution from the surface tension of the liquid at the contact line on the particle surface, the second term captures the contribution of the Laplace pressure with the gravitational effect included, and the last term is a buoyancy force.

The full Eq.~(\ref{eq:Young-Laplace}) is hard to solve analytically. The main difficulty is the presence of the gravitational term, the importance of which is captured by a capillary length defined as $\kappa^{-1} = \sqrt{\frac{\gamma}{\Delta \rho g}}$. For water at room temperature, $\kappa^{-1} \approx 2.7$ mm. In this paper, we are mainly concerned about a small particle with size ranging from nanometer to micrometer scales. Therefore, it is always the case that $R \ll \kappa^{-1}$. In the limit of $R< L \ll \kappa^{-1} $, the gravitational term in Eq.~(\ref{eq:Young-Laplace}) is negligible and the equation can be solved analytically with the elliptic integrals. In the opposite limit $R \ll \kappa^{-1} \ll L$, the approximate Derjaguin-James formula exists which can be used to estimate $\Delta h$ \citep{Derjaguin1946, James1974}. In the crossover region where $L \sim \kappa^{-1}$, we recently found another approximate formula \cite{Tang2018arXiv}, based on numerical solutions of Eq.~(\ref{eq:Young-Laplace}), to predict $\Delta h$. In the following we first discuss a way to transform Eq.~(\ref{eq:Young-Laplace}) that allows numerical treatments in general and then analyze the different regions in detail.

Eq.~(\ref{eq:Young-Laplace}) can be made dimensionless via a variable change,
\begin{align}
x \equiv \kappa r~,\quad y \equiv \kappa z_0~.
\end{align}
The result is the following nonlinear differential equation for a rising meniscus,
\begin{align} \label{eq:YL_dimensionless}
\frac{x''}{(1 + x'^2)^{3/2}} - \frac{1}{x(1 + x'^2)^{1/2}} 
= \frac{2 \tilde{H}}{\kappa} + y~, 
\end{align}
where $x' \equiv \frac{\mathrm{d}x}{\mathrm{d}y}$, $x'' \equiv \frac{\mathrm{d}^2x}{\mathrm{d}y^2}$, and $2 \tilde{H} \equiv \Delta p/\gamma$ is the local mean curvature of the meniscus. Again, the two terms on the left hand side need to flip signs for a depressed meniscus. This second-order differential equation can be rewritten into two coupled first-order differential equations for which numerical treatments are much easier. To this end, we take the local normal vector of the liquid-vapor interface pointing toward the liquid phase and introduce an angle parameter $t$, which is the angle of rotating the $z$-axis \textit{clockwise} to the local normal vector, as shown in Fig.~\ref{fg:main}. It is easy to show that $t$ always changes from $\pi$ at the surface of the container to $\theta_1 + \psi$ on the particle surface. For a rising meniscus, $\theta_1 + \psi < \pi$  while for a depressed one, $\theta_1 + \psi > \pi$.

Eq.~(\ref{eq:YL_dimensionless}) can be rewritten in terms of $t$ as
\begin{align} \label{eq:YL_dimensionless_2}
\frac{\mathrm{d}\sin t}{\mathrm{d}x} + \frac{\sin t}{x} = - \frac{2 \tilde{H}}{\kappa} - y~.
\end{align}
Eq. (\ref{eq:YL_dimensionless_2}) is actually more general than Eq.~(\ref{eq:YL_dimensionless}) as the former applies no matter the meniscus is rising or depressed while the latter only describes a rising meniscus, though both can deal with a meniscus with a neck. It is thus advantageous to use Eq. (\ref{eq:YL_dimensionless_2}) to describe an (axisymmetric) meniscus on the outside of a particle at a liquid-vapor interface.

Rewriting Eq.~(\ref{eq:YL_dimensionless_2}) for $\frac{\mathrm{d}x}{\mathrm{d}t}$ and using $\frac{\mathrm{d}y}{\mathrm{d}x} = \tan t$, we obtain a pair of coupled first-order nonlinear differential equations,
\begin{subequations}  
  \begin{align}
	\frac{\mathrm{d}x}{\mathrm{d}t} 
	= - \left(\frac{2\tilde{H}}{\kappa} + y + \frac{\sin t}{x}\right)^{-1} 
	\cos t~, \\	
	 \frac{\mathrm{d}y}{\mathrm{d}t} 
	= - \left(\frac{2\tilde{H}}{\kappa} + y + \frac{\sin t}{x}\right)^{-1} 
	\sin t~,
  \end{align}
  \label{eq:YL_1st_order}
\end{subequations}
with the following boundary conditions,
\begin{subequations}  
  \begin{align}
	t &= \theta_1 + \psi \quad \text{at} \quad x = \kappa R\sin \psi~, \label{eq:BC_1st_order_a} \\
	t &= \pi \quad \text{at} \quad x = \kappa L \quad \text{and}  \quad y = 0~. \label{eq:BC_1st_order_b} 
  \end{align}
\end{subequations}
Generally, numerical solutions of Eq.~(\ref{eq:YL_1st_order}) can be obtained by the shooting method \citep{Press2007}. It should be emphasized that Eq.~(\ref{eq:YL_1st_order}) provides a unified description for a rising or depressed meniscus with or without a neck. The difference only shows up in the boundary condition in Eq.~(\ref{eq:BC_1st_order_a}). For a meniscus rising on the particle surface, $\theta_1 + \psi < \pi$ while for a depressed one, $\theta_1 + \psi > \pi$. Moreover, if $\theta_1 + \psi < \pi/2$ or $\theta_1 + \psi > 3\pi/2$, then the meniscus has a neck.

\subsection{The Limit $L \ll \kappa^{-1} $} \label{sec:2b}

In the limit of $R< L \ll \kappa^{-1} $, the Bond number ${\rm Bo} \equiv gL^2\Delta \rho/\gamma = \kappa^2L^2$ is much smaller than 1, indicating that gravity can be ignored in the treatment of the meniscus. Eq.~(\ref{eq:Young-Laplace}) for a rising meniscus can be simplified as
\begin{align}
\frac{r''}{(1 + r'^2)^{3/2}} - \frac{1}{r (1 + r'^2)^{1/2}} = 2 \tilde{H}
\label{eq:Young-Laplace-no-g}
\end{align}
with $2 \tilde{H} \equiv \Delta p/\gamma$. For a depressed meniscus, the right hand side of Eq.~(\ref{eq:Young-Laplace-no-g}) should be $-2 \tilde{H}$. Eq.~(\ref{eq:Young-Laplace-no-g}) can be solved analytically for the boundary condition sketched in Fig. \ref{fg:main}. The derivation below benefits from a seminal paper of Orr \textit{et al.} on the theory of pendular rings \cite{Orr1975} and a recent work by Rubinstein and Fel \citep{Rubinstein2014}.

We first make Eq.~(\ref{eq:Young-Laplace-no-g}) dimensionless by introducing new variables $X = r/R$ and $Y= z_0/R$. Defining $u= \sin t$ with the angle parameter $t$ introduced previously, we can rewrite Eq.~(\ref{eq:Young-Laplace-no-g}) as
\begin{align} \label{eq:yl_reduce}
-2H = \frac{\mathrm{d}u}{\mathrm{d}X} + \frac{u}{X}~,
\end{align}
where $H \equiv R\tilde{H}$ is the dimensionless mean curvature of the meniscus. The boundary conditions are
\begin{subequations}  
  \begin{align}
	t &= t_1 \quad \text{at} \quad X_1 = \sin \psi~, \label{eq:b1a} \\
	t &= t_2\quad \text{and}\quad Y_2 = 0 \quad \text{at} \quad X_2 = l~, \label{eq:b1b} 
  \end{align}
\end{subequations}
where $t_1 = \theta_1 + \psi$, $t_2=\pi$, and $l = L/R > 1$ is the scaled radius of the bucket. Like Eq.~(\ref{eq:YL_dimensionless_2}), Eq.~(\ref{eq:yl_reduce}) applies to both rising and depressed menisci and is more general than Eq.~(\ref{eq:Young-Laplace-no-g}).

The solution for Eq. (\ref{eq:yl_reduce})
is 
\begin{align}
u = \frac{c}{4HX} - HX~. 
\label{eq:solution}
\end{align}
The boundary condition Eq.(\ref{eq:b1a}) indicates that
\begin{align}
c = 4H \sin \psi \left[ H \sin \psi + \sin (\theta_1 + \psi) \right]~.
\end{align}
The other boundary condition Eq.(\ref{eq:b1b}) yields
\begin{align}
c = 4 H^2 l^2~.
\end{align}
As a result, $H$ is given by
\begin{align}
H = \frac{\sin \psi \sin (\theta_1 + \psi)}{l^2 - \sin^2 \psi}~.
\label{eq:H}
\end{align}
Note that $l \equiv L/R > 1$ and therefore the denominator in Eq.~(\ref{eq:H}), $l^2 - \sin^2 \psi$, is always positive. For a rising meniscus, $H>0$ as $0<\theta_1 + \psi < \pi$ while for a depressed one, $H<0$ as $\pi< \theta_1 + \psi <2 \pi$. When the particle is in its equilibrium location, $\theta_1 + \psi = \pi$ and $H=0$, as expected for a flat liquid-vapor interface. Generally, $H$ asymptotically approaches 0 when $l \equiv L/R \rightarrow \infty$.

Equation (\ref{eq:solution}) yields a parametric relation, $X(t)$, which must be positive definite. Then the solution of the meniscus profile, $Y(t)$, can be determined by noting that $\text{d}Y/\text{d}X = \tan t$. The results are
\begin{align}
X(t) &= \frac{1}{2H}\left(-\sin t \pm \sqrt{\sin^2 t + c}~\right),   \label{eq:xt} \\
Y(t) &= \frac{1}{2H} \int_{t_2}^{t} \left(-\sin \phi \pm 
\frac{\sin^2 \phi}{\sqrt{\sin^2 \phi +c}}\right) \, \mathrm{d}\phi~, \label{eq:yt}
\end{align}
where the $+$ ($-$) sign is for a rising (depressed) meniscus. Hereafter, when the sign $\pm$ or $\mp$ appears in an equation, the upper sign is always for a rising meniscus while the lower sign is for a depressed one. In general, $t_2 = 3\pi/2 - \theta_2$ where $\theta_2$ is the contact angle of the liquid on the container surface. For the systems considered here, $\theta_2 = \pi/2$ and therefore $t_2 = \pi$.

Equation (\ref{eq:yt}) can be evaluated by the elliptic integrals and the result is
\begin{align} \label{eq:yt_elliptic}
Y(t) &= \frac{1}{2H}(\cos t - \cos t_2)  \nonumber \\
&\pm \frac{\sqrt{c}}{2H} \Big[ E(t,j) - E(t_2,j) - F(t,j) + F(t_2, j) \Big]~,
\end{align}
where $j^2 \equiv -\frac{1}{c}$, $E(t, j) \equiv \int_0^{t} \sqrt{1 - j^2 \sin^2\phi} \, \mathrm{d}\phi$ is the incomplete elliptic integral of the second kind, and $F(t, j) \equiv \int_0^t \frac{1}{\sqrt{1 - j^2 \sin^2 \phi}} \, \mathrm{d} \phi$ is the incomplete elliptic integral of the first kind, respectively. The meniscus height, $\Delta h$, is given by $Y(t_1)$ with $t_1 = \theta_1 + \psi$ and the result is
\begin{align}\label{eq:height_elliptic}
\Delta h & = \frac{R}{2H}\Big[ \cos ( \theta_1+\psi) + 1 \Big] \nonumber \\
&\pm \frac{R \sqrt{c}}{2H} 
\Big[ E(\theta_1+\psi -\pi, j) - F(\theta_1 + \psi-\pi, j)\Big]~.
\end{align}

Equation (\ref{eq:height_elliptic}) holds as long as $\kappa^{-1} \gg L > R$. An approximate formula can be derived for $\Delta h$ in the limit of $\kappa^{-1} \gg L \gg R$ using the series expansions of the elliptic integrals and the asymptotic behavior $H=\frac{\sin \psi \sin (\theta_1 + \psi)}{l^2 - \sin^2 \psi} \simeq \frac{\sin \psi \sin (\theta_1 + \psi)}{l^2} \rightarrow 0$. The result is
\begin{align} \label{eq:height_logL}
\Delta h &\simeq R \sin \psi \sin ( \theta_1 + \psi ) \nonumber \\
&\times \Big\{  \ln \frac{2L}{ R \sin \psi [1 - \cos (\theta_1+\psi)]} - \frac{1}{2} \Big\}~,
\end{align}
which indicates that $\Delta h \sim R \ln (L/R)$ for $\kappa^{-1} \gg L \gg R$. Our numerical results indicate that this scaling relationship holds up to about $L \lesssim 0.4\kappa^{-1}$ \cite{Tang2018arXiv}. Eq.~(\ref{eq:height_logL}) is very close to the result for a catenoid for which $H=0$, except for the $-1/2$ term in the curly bracket \cite{Tang2018arXiv}.

Again, a self-consistent procedure using the constraint that the volume of the liquid bath is fixed needs to be employed to determine the filling angle $\psi$ for a given displacement, $\Delta z$. This procedure can be facilitated if we note that for the solution of the meniscus profile given in Eqs.~(\ref{eq:xt}) and (\ref{eq:yt_elliptic}), the volume of the liquid bath can be expressed analytically as
\begin{align}
\frac{V}{R^3} = \frac{\pi}{8H^3} J_s + \pi l^2 \frac{h}{R}
- \frac{1}{3}\pi(2 - 3 \cos \psi + \cos^3 \psi)~, 
\end{align} \label{eq:vol}
where 
\begin{align}
J_s &= (4+c)(\cos t_1 - \cos t_2) \nonumber \\
&- \frac{4}{3}\left(\cos^3 t_1 - \cos^3 t_2\right) \nonumber \\
&\pm\sqrt{c} \left\{ \frac{8+c}{3} \Big[ E(t_1, k) - E(t_2, k) \Big] \right.  \nonumber \\
&\left. - \frac{4+c}{3} \Big[ F(t_1,k) -F(t_2, k) \Big] \right\} \nonumber \\
&\mp \frac{2}{3}\left[ \sin (2t_1) \sqrt{\sin^2 t_1 +c } -\sin (2t_2) \sqrt{\sin^2 t_2 +c}~\right]~.
\end{align}

After $\psi$ is determined for a given $\Delta z$, the capillary force in the limit of $\kappa^{-1} \gg L > R$ can be computed as
\begin{align} \label{eq:force_H}
F = 2\pi\gamma R \Big[ \sin \psi \sin(\theta_1 + \psi) - H \sin^2 \psi \Big]
\end{align}
with $H$ given in Eq.~(\ref{eq:H}).

Since for $l = L/R \gg 1$, the dimensionless mean curvature $H\rightarrow 0$, the capillary force is dominated by the surface tension term,
\begin{align} \label{eq:force_approx}
F \simeq 2\pi\gamma R \sin \psi \sin(\theta_1 + \psi)~,
\end{align}
and the meniscus height can be approximated as
\begin{align} \label{eq:meniscus_rise_approx}
\Delta h \simeq R \sin \psi \sin ( \theta_1 + \psi ) \ln \frac{2L}{R}~.
\end{align}
From Eqs.~(\ref{eq:D}) and (\ref{eq:interface_height_container}), we have
\begin{align} \label{eq:delta_z}
\Delta z = \Delta h + h - h_0 + R (\cos\theta_1 + \cos\psi)~.
\end{align}
For $l = L/R \gg 1$, the meniscus height satisfies $\Delta h \gg R(\cos \theta_1 + \cos \psi)$ and the displacement of the liquid-vapor interface far away from the particle becomes negligible, i.e., $h= h_0$. Therefore, we have approximately
\begin{align} \label{eq:delta_z_delta_h}
\Delta z \simeq \Delta h~.
\end{align}
Combining Eqs.~(\ref{eq:force_approx}), (\ref{eq:meniscus_rise_approx}), and (\ref{eq:delta_z_delta_h}), we finally arrive at the Joanny-de Gennes' Hookean law \cite{Joanny1984}, 
\begin{align} \label{eq:deGennes-Hooke}
F \simeq \frac{2 \pi \gamma}{\ln(2L/R)} \Delta z~,
\end{align}
yielding an effective spring constant for a particle at a liquid-vapor interface that softens with the lateral span of the interface as $\ln^{-1} (2L/R)$. The denominator $\ln (2L/R)$ in Eq.~(\ref{eq:deGennes-Hooke}) was absent in the form derived by Pieranski \cite{Pieranski1980} and used in many work including the recent ones on modeling solvent evaporation with the moving interface method \cite{Fortini2016,Fortini2017,Howard2017,Reinhart2017}. This omission is easy to understand as in Pieranski's model, the liquid-vapor interface is always flat even for a particle out of its equilibrium location at the interface \cite{Pieranski1980}. However, in the model discussed here the meniscus height on the outside of the particle scales with $\ln (2L/R)$.

\subsection{The region $L \gtrsim \kappa^{-1} $} \label{sec:2c}

Our previous work indicated that the theory presented in Sec.~\ref{sec:2b} can be used to describe a meniscus on the outside of a small circular cylinder pretty accurately for $L$ up to $\sim 0.4\kappa^{-1}$ \cite{Tang2018arXiv}. For $L \gtrsim 4\kappa^{-1}$, the interface can be treated as unbound and the meniscus height is given by the Derjaguin-James formula
\begin{align} \label{eq:height_DJ}
\Delta h &= R \sin \psi \sin ( \theta_1 + \psi ) \nonumber \\
&\times \Big\{  \ln \frac{4\kappa^{-1}}{ R \sin \psi [1 - \cos (\theta_1+\psi)]} - E \Big\}~,
\end{align}
where $E = 0.57721...$ is the Euler-Mascheroni constant.

In the crossover region $0.4\kappa^{-1} \lesssim L \lesssim 4\kappa^{-1}$, our previous work revealed an approximate expression of $\Delta h$ as \cite{Tang2018arXiv}
\begin{align}
\Delta h &= \Delta h (\textrm{elliptic})  \nonumber \\
& \times \Big\{1-m(\kappa L) \Big(\kappa R \sin \psi [1-\cos (\theta_1+\psi)] \Big)^{0.12}\Big\}~,
\label{eq:height_final}
\end{align}
where $\Delta h (\textrm{elliptic})$ is the expression of the meniscus height in Eq.~(\ref{eq:height_elliptic}) based on the elliptic integrals and $m(x)$ is a kink function that reads
\begin{equation}
m(x) = \begin{cases}
0.085 \exp{\left[ (x-1.85)^{1.83}/0.74 \right]} &\text{if $x \le 1.85$~,}\\
0.085 \exp{\left[ (1.85-x)/0.875 \right]} &\text{if $x > 1.85$~.}
\end{cases}
\label{eq:fit_kink_function}
\end{equation}

Noting that $l \equiv L/R$ is the key parameter entering the expression of $\Delta h (\textrm{elliptic})$ in Eq.~(\ref{eq:height_elliptic}). In the Derjaguin-James formula [Eq.~(\ref{eq:height_DJ})] the term $L/R$ is replaced by $\kappa^{-1}/R$. We can therefore make Eq.~(\ref{eq:height_final}) applicable for any arbitrary $L$ (as long as it is larger than $R$) by using the following definition of $l$ for $\Delta h (\textrm{elliptic})$,
\begin{equation}
l = \begin{cases}
L/R &\text{if $L \le 1.85\kappa^{-1}$~,}\\
1.85\kappa^{-1}/R &\text{if $L > 1.85\kappa^{-1}$~.}
\end{cases}
\label{eq:l_parameter}
\end{equation}
The particular choice of the cutoff, $1.85\kappa^{-1}$, can be understood by equating $\Delta h$ in Eq.~(\ref{eq:height_logL}), which is a close approximation of Eq.~(\ref{eq:height_elliptic}), to that in the Derjaguin-James formula in Eq.~(\ref{eq:height_DJ}). At $L = 2 e^{1/2 -E}\kappa^{-1} \approx 1.85\kappa^{-1}$, the two expressions are equal. With $l$ defined in Eq.~(\ref{eq:l_parameter}), the meniscus height $\Delta h$ in Eq.~(\ref{eq:height_final}) reduces to the expression in Eq.~(\ref{eq:height_elliptic}) for $L \ll \kappa^{-1}$, while to the Derjaguin-James formula in Eq.~(\ref{eq:height_DJ}) for $L \gg \kappa^{-1}$. It also agrees well with the numerical solutions of $\Delta h$ in the crossover region where $L \sim \kappa^{-1}$ \cite{Tang2018arXiv}.

For $L \gtrsim \kappa^{-1}$, the displacement of the liquid-vapor interface (i.e., $h-h_0$) induced by the particle displacement, $\Delta z$, is negligible and thus $h \simeq h_0$. As a result
\begin{align} \label{eq:delta_z_largeL}
\Delta z \approx \Delta h  + R (\cos\theta_1 + \cos\psi)~.
\end{align}
Eq.~(\ref{eq:delta_z_largeL}) and Eq.~(\ref{eq:height_final}) together provide an expression of $\Delta z$ with the filling angle $\psi$ as a parameter for $L \gtrsim \kappa^{-1}$ (other physical quantities, $L$, $R$, and $\theta_1$ are already known when the problem is set-up). In this limit, there is no need to use the volume constraint of the liquid bath to connect $\psi$ to $\Delta z$.

The capillary force in the region $L \gtrsim \kappa^{-1}$ becomes
\begin{align} \label{eq:force_approx_2}
F &= 2\pi\gamma R \Big[ \sin \psi \sin(\theta_1 + \psi) - \frac{1}{2}\kappa^2 \Delta h R \sin^2 \psi   \nonumber \\
&- \frac{1}{6} \kappa^2R^2 (2 - 3\cos \psi + \cos^3 \psi) \Big]~.
\end{align}
The Laplace pressure term drops out because $H \simeq 0$ in this limit. In this paper we are concerned about particles with $R \ll \kappa^{-1}$ and the buoyancy force and the gravitational term above are negligible compared to the surface tension contribution. As a result, the capillary force is dominated by the surface tension term. Eq.~(\ref{eq:force_approx_2}) provides an expression of the capillary force with $\psi$ as a parameter. Combining Eqs.~(\ref{eq:force_approx_2}), (\ref{eq:delta_z_largeL}), and (\ref{eq:height_final}), we obtain a force-displacement curve parameterized by $\psi$ for $L \gtrsim \kappa^{-1}$.

\section{Simulation Methods} \label{sec:simu}

A snapshot from MD simulations of a particle at a liquid-vapor interface is shown in Fig.~\ref{fg:main}(b). In order to address generic behavior, we consider a molecular liquid consisting of short linear chains of four spherical beads. This tetramer model captures many aspects of the behavior of hydrocarbon chains \cite{Cheng2014,Cheng2016Langmuir}. All the beads interact with a Lennard-Jones (LJ) potential,
\begin{align}
V_{\text{LJ}}(a) = 4\epsilon \left[ \left( \frac{\sigma}{a} \right)^{12} 
-\left(\frac{\sigma}{a}\right)^6 - \left(\frac{\sigma}{a_c}\right)^{12}
+ \left(\frac{\sigma}{a_c}\right)^6 \right]~,
\end{align}
where $a$ is the distance between the centers of beads, $\sigma$ represents an effective bead diameter, and $\epsilon$ is an energy scale. The LJ potential is truncated at $a_c = 2.5\sigma$. Two neighboring beads on a chain are connected by a bond described by a finitely extensible nonlinear elastic (FENE) potential \cite{Kremer1990}, 
\begin{align}
V_{\text{FENE}}(a) = -\frac{1}{2} K R_0^2 \ln \left( 1- \frac{a^2}{R_0^2}\right)~,
\end{align}
where the canonical values are adopted with $R_0 = 1.5 \sigma$ and $K = 30 \epsilon/\sigma^2$.

A spherical particle is modeled as a uniform distribution of LJ mass points. The interaction between the particle and a LJ bead is determined by integrating the LJ potential between the bead and all the mass points on the particle~\cite{Everaers2003, IntVeld2008}. The resulting potential is
\begin{align}
U_{\text{ns}}(a) &= \frac{2}{9}\frac{R_\text{n}^3 \sigma^3 A_{\text{ns}}}
{(R_\text{n}^2 - a^2)^3} \nonumber \\
&\times \left[ 1 - \frac{(5R_\text{n}^6 + 45R_\text{n}^4 a^2 + 63 R_\text{n}^2 a^4+15 a^6)\sigma^6}
{15(R_\text{n}-a)^6 (R_\text{n}+a)^6} \right]~, 
\end{align}
where $a$ is the center-to-center distance between the bead and particle and the radius of the particle is $R_{\text{n}} = 10\sigma$ in our simulations. If we take $\sigma \sim 0.5$nm, then $R_\text{n}$ is about 5nm. The potential is truncated at $a_c = 14\sigma$. The Hamaker constant $A_{\text{ns}}$ controls the wetting behavior of the liquid on the particle surface.

The liquid bath is placed in a cylindrical container of nominal radius $L_\text{n}$. Two values, $L_{\text{n}} = 50 \sigma$ and $ 75 \sigma$, are used in simulations. The central axis of the container is along the $z$-axis. The liquid-vapor interface is flat in the horizontal $x$-$y$ plane when the particle is in equilibrium at the interface. Two horizontal walls are used at $z=0$ and $z=100\sigma$ to confine the liquid and vapor. The interaction between a LJ bead and the top, bottom, or side wall is governed by a LJ 9-3 potential
\begin{align}
U_{\text{w}}(a) = \epsilon_{\text{w}} \left[ \frac{2}{15}\left( \frac{\sigma}{a}\right)^9 
- \left( \frac{\sigma}{a} \right)^3 
- \frac{2}{15} \left( \frac{\sigma}{a_\text{c}}\right)^9
+ \left( \frac{\sigma}{a_\text{c}} \right)^3 \right]~,
\end{align}
where $a$ is the distance from the bead center to the wall and $a_\text{c} = 2.5\sigma$, $0.8583\sigma$, and $3.0\sigma$ are the cutoff distances at the side, top, and bottom wall, respectively. The interaction strength is set with $\epsilon_{\text{w}} = 2.1\epsilon$ to yield a contact angle $\sim 90^\circ$ on the side wall, as confirmed with independent simulations.

The tetramer LJ liquid has density $\rho_l = 0.927 m/\sigma^3$ and vapor density $\rho_v=0$ for the temperature used in simulations. Our motivation to pick this nonvolatile liquid is to generate a liquid-vapor interface that is sharp and can equilibrate quickly. To determine the surface tension of the tetrameter liquid, we simulated a liquid film in a cubic simulation cell with a liquid-vapor interface in the $x$-$y$ plane, in which the periodic boundary conditions are used. The liquid film is in contact with a bottom wall at $z=0$ and confined by a top wall at $z=100\sigma$. The liquid-vapor interface is at about $z=51\sigma$. The surface tension was computed with the Kirkwood-Buff formula~\cite{Kirkwood1949},
\begin{align}
\gamma = \frac{1}{2} \int \left[ p_{zz}(z) - \frac{p_{xx}(z) + p_{yy}(z)}{2} \right] \, \mathrm{d}z~,
\end{align}
where $p_{xx}(z)$, $p_{yy}(z)$, and $p_{zz}(z)$ are the three diagonal components of the stress tensor. The result is $\gamma = 1.018 \epsilon/\sigma^2$ in the LJ units. A rough mapping to real units can be found in a previous study \cite{Cheng2014}.

\begin{figure}[tp]
  \includegraphics[width = 0.4 \textwidth ]{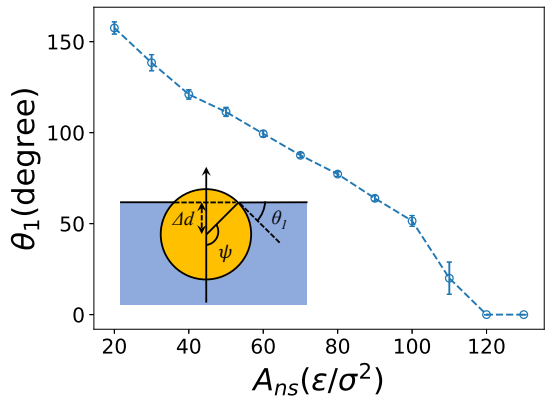}
  \caption{Contact angle $\theta_1$ vs. Hamaker constant $A_{\text{ns}}$ for a particle with $R_\text{n} = 10\sigma$ at the surface of the tetramer liquid. The inset shows the equilibrium configuration of the particle straddling the liquid-vapor interface, where $\psi = \pi-\theta_1$.}
  \label{fg:contact_angle}
\end{figure}

The wetting behavior of the tetramer liquid on the particle surface depends on the Hamaker constant $A_{\text{ns}}$. Considering the finite radius of the particle, we identified the contact angle of the liquid on the particle surface directly by placing the particle at the surface of the liquid film that was used in the calculation of surface tension. When the particle settles into its equilibrium location, the liquid-vapor interface is flat and intersects with the particle surface with a filling angle $\psi = \pi-\theta_1$. The particle center is then at distance $\Delta d = R\cos \theta_1$ below the liquid-vapor interface. By computing $\Delta d$ in MD simulations, we can determine the contact angle as
\begin{align}
\theta_1 = \arccos(\Delta d/R)~.
\end{align}
To determine $\Delta d$, we need to obtain the location of the liquid-vapor interface, which was achieved by fitting the density profile of the liquid away from the location of the particle to the following functional form,
\begin{align} \label{eq:interface}
\rho(z) = \frac{1}{2}(\rho_l + \rho_v) - \frac{1}{2} (\rho_l - \rho_v)
\tanh \left[ \frac{2(z - z_i)}{d_s} \right]~,
\end{align} 
where $z_i$ is the location and $d_s$ is the width of the interface, respectively.

The results for $\theta_1$ as a function of $A_{\text{ns}}$ are shown in Fig.~\ref{fg:contact_angle}. The contact angle decreases when the Hamaker constant between the particle and liquid increases \cite{Cheng2012, Koplik2017}. This trend is expected as stronger interactions between a solid surface and a liquid favor the wetting of the solid by the liquid. The contact angle $\theta_1$ and surface tension $\gamma$ are used as material properties when the simulation results of the meniscus on the outside of a particle are compared to the predictions of the theory of capillarity. 

We used a pulling process to place the particle at various locations along the vertical $z$-axis across the liquid-vapor interface. First, the particle was fully immersed in the liquid bath. Then the particle was pulled upward with a constant speed $v = 0.02 \sigma/\tau \sim 4$ m/s along the $z$ direction. This speed is 6 orders of magnitude larger than typical velocities of displacing particles ($\sim 1~\mu$m/s) in AFM experiments \cite{Anachkov2016}. To get rid of the inertial effects \cite{Cheng2014}, we allowed the meniscus to relax for at least $5000 \tau$ when the particle was pulled to and fixed at a certain location. After relaxation the meniscus profile was determined from the density profile of the liquid and the capillary force on the particle was computed. The procedure was repeated when the particle was pulled to its next location until the meniscus broke up. The parameters of all five systems studied with MD simulations are summarized in Table \ref{tb:table1}.
\begin{table}
\centering
\caption{Parameters of all five systems studied in this paper.}
\label{tb:table1}
\begin{tabular}{llllll}
\hline
$A_{\text{ns}}~~~$  & $\theta_1~~~$  & $R_{\text{n}}/\sigma~~~$ & $L_{\text{n}}/\sigma~~~$ & $h_0/\sigma~~~$ & \# of tetramers\\ \hline
100               & $48.5^\circ$      & 10    &  50        & 51.4  & 90000\\
80                & $76.5^\circ$       & 10    &  50       & 51.3  & 90000\\
60                & $98.2^\circ$       & 10    &  50      & 51.1  & 90000\\
40                & $120.5^\circ$     & 10    &  50        & 51.0 & 90000 \\
100               & $48.5^\circ$      & 10    &  75       & 51.1  & 202612 \\ \hline
\end{tabular}
\end{table}

The large-scale atomic/molecular massively parallel simulator (LAMMPS) developed at Sandia National Laboratories~\cite{Plimpton1995} was adopted to perform all the MD simulations reported here. A velocity-Verlet algorithm was used to integrate the equation of motion with a time step $\delta t = 0.005 \tau$, where  $\tau = \sigma (m/\epsilon)^{1/2}$ is the LJ unit of time and $m$ is the mass of a LJ bead. The particle has mass $M = \frac{4\pi R^3 m}{3\sigma^3} = 4188.79m$. In all the simulations, the liquid was held at $T=0.7 \epsilon/k_{\text{B}}$ via a Langevin thermostat with a damping rate $\Gamma = 0.1 \tau^{-1}$.

\section{Results and Discussion} \label{sec:RD}

\subsection{Theoretical procedure to determine filling angle} \label{ss:procedure_filling_angle}

\begin{figure}[ht]
  \includegraphics[width = 0.35 \textwidth]{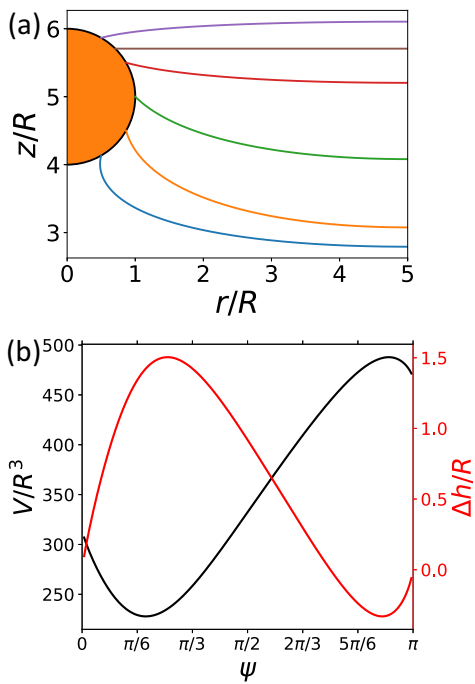}
  \caption{(a) Meniscus profiles for $\theta_1 = \pi/4$ and different filling angles: $\psi = \pi/6$, $\pi/3$, $\pi/2$, $2\pi/3$, $3\pi/4$, $5\pi/6$ (from bottom to top). The ratio $L/R = 5$ is used here and the center of the particle is fixed at $z = 5R$ (i.e., $D=4R$ in Fig.~\ref{fg:main}). (b) The volume under the meniscus profile (measured from the bottom wall at $z=0$) and the height of the meniscus (measured from the contact line on the container surface) as a function of $\psi$ for the parameters in (a).}
  \label{fg:numerical}
\end{figure}

With the theory in Sec.~\ref{sec:2b}, the meniscus profile can be predicted for the region $L \ll \kappa^{-1}$ when the contact angle $\theta_1$ and the filling angle $\psi$ on the particle surface are given. Some examples for $\theta_1 = \pi/4$, $L/R = 5$, $D=4R$ (i.e., with the particle center fixed at $z=5R$), and various values of $\psi$ are given in Fig.~\ref{fg:numerical}(a). The corresponding volume under the meniscus profile ($V$) and the height of the meniscus rise or depression ($\Delta h$) are shown in Fig.~\ref{fg:numerical}(b). It should be pointed out that the meniscus profile, $r(z_0)$, depends on $R$, $L$, $\theta_1$, and $\psi$, but not on $D$. However, as discussed in Sec.~\ref{sec:2b} for the setup in Fig.~\ref{fg:main}, the volume of the liquid bath is conserved. When the particle is pulled or pushed vertically to a certain height, the filling angle of the meniscus that is physically realized needs to satisfy the volume constraint. This fact is easy to understand as $D$, which sets the vertical location of the particle, is the parameter controlled in both simulations and experiments. The filling angle $\psi$ is then a parameter set by the fixed volume of the liquid bath.

As shown in Fig.~\ref{fg:numerical}(b), $V$ and $\Delta h$ are anticorrelated but nonmonotonic when $\psi$ varies. This behavior indicates that for a certain range of initial volumes of the liquid bath, there could be two possible values of $\psi$ satisfying the volume constraint. To determine which $\psi$ is the physical solution, we will only consider filling angles $\psi \in [\psi_{\text{min}}, \psi_{\text{max}}]$ with $\psi_{\text{min}}$ and $\psi_{\text{max}}$ being the solutions of $\frac{\mathrm{d} V}{\mathrm{d} \psi} = 0$. In particular, $V$ has a minimal value $V_{\text{min}}$ at $\psi = \psi_{\text{min}}$ and a maximal value $V_{\text{max}}$ at $\psi=\psi_{\text{max}}$. Clearly, $V$ is a monotonically increasing function of $\psi \in [\psi_{\text{min}}, \psi_{\text{max}}]$. Then the initial volume of the liquid bath, which is set by $R$, $L$, $\theta_1$, and $D_0$ or $h_0$ through Eq.~(\ref{eq:vol_initial}), can be used to determine the filling angle $\psi$ for the given $D$, which sets the displacement of the particle, $\Delta z$, through Eq.~(\ref{eq:D}). With $\psi$ determined, the meniscus profile and the capillary force on the particle can be readily computed. If the initial volume of the liquid bath is smaller than $V_{\text{min}}$ at the given $D$, then the meniscus is assumed to be ruptured and the particle is completely in the vapor. On the contrary, if the initial volume is larger than $V_{\text{max}}$ at the given $D$, then the particle is fully immersed in the liquid bath.

We can use intuitive arguments to justify the criterion adopted here of only picking $\psi\in [\psi_{\text{min}}, \psi_{\text{max}}]$ for a given $D$. For a particle in its equilibrium height at a liquid-vapor interface, the interface is flat and the filling angle $\psi = \pi-\theta_1$. When the particle is pulled upward, the filling angle starts to decrease from $\pi-\theta_1$ as the contact line is free to slide. It is natural to assume that the filling angle changes continuously when the particle is pulled higher and higher until at a critical filling angle the meniscus ruptures and detaches from the particle surface. On the contrary, if the particle is pushed downward from its equilibrium height, then the filling angle increases continuously from $\pi-\theta_1$ until at another critical filling angle the meniscus collapses and the particle submerges into the liquid. In any case, a filling angle further away from 0 (for the case of the particle being pulled upward from its equilibrium location) or $\pi$ (for the case of pushing downward) is more physically possible than the other one when there are two possible filling angles that satisfy the volume constraint. Therefore, it is physically sensible to exclude filling angles less than $\psi_{\text{min}}$ and those larger than $\psi_{\text{max}}$.

Confining $\psi\in [\psi_{\text{min}}, \psi_{\text{max}}]$ can also be justified mathematically. First we consider the case where $D$ is fixed as in Fig.~\ref{fg:numerical}. Note that $h$ and $\Delta h$ are still functions of $\psi$. Using $\mathrm{d} D = 0$ and Eq.~(\ref{eq:interface_height_container}), we obtain
\begin{align} \label{eq:derivative_constraint}
\frac{\mathrm{d} h}{\mathrm{d} \psi} + \frac{\mathrm{d} \Delta h}{\mathrm{d} \psi} - R \sin \psi= 0~,
\end{align} 
which is a constraint that the derivatives of $h$ and $\Delta h$ have to satisfy. In this case, $V$ is a function of $\psi$ through Eq.~(\ref{eq:vol_bath}), from which we get
\begin{align} \label{eq:vol_derivative}
\frac{\mathrm{d} V}{\mathrm{d} \psi} = \pi [r(\Delta h)]^2 \frac{\mathrm{d} \Delta h}{\mathrm{d} \psi} + \pi L^2\frac{\mathrm{d} h}{\mathrm{d} \psi} - \pi R^3 \sin^3 \psi~.
\end{align} 
Using $r(\Delta h)= R\sin\psi$ and Eq.~(\ref{eq:derivative_constraint}), we can rewrite $\frac{\mathrm{d} V}{\mathrm{d} \psi}$ as
\begin{align} \label{eq:vol_derivative_2}
\frac{\mathrm{d} V}{\mathrm{d} \psi} = \pi \left( R^2 \sin^2\psi - L^2\right)
\left(\frac{\mathrm{d} \Delta h}{\mathrm{d} \psi}  - R \sin \psi \right)~.
\end{align} 
Therefore, the condition $\frac{\mathrm{d} V}{\mathrm{d} \psi} = 0$ when $D$ is fixed is equivalent to
\begin{align} \label{eq:meniscus_height_derivative}
\frac{\mathrm{d} \Delta h}{\mathrm{d} \psi}  - R \sin \psi = 0~.
\end{align} 
The boundary value of the filling angle, $\psi_{\text{min}}$ and $\psi_{\text{max}}$, can be determined using Eq.~(\ref{eq:meniscus_height_derivative}). Since the meniscus profile, $r(z_0)$, does not depend on $D$. Eq.~(\ref{eq:meniscus_height_derivative}) is independent of $D$, indicating that $\psi_{\text{min}}$ and $\psi_{\text{max}}$ are $D$-independent as well.

Eq.~(\ref{eq:meniscus_height_derivative}) can also be understood from a different perspective based on the intuitive arguments discussed previously. Now we consider a liquid bath with a fixed volume $V$. When the particle is in its equilibrium height at the surface of the liquid bath, the filling angle is $\pi - \theta_1$. When the particle is pulled upward (or pushed downward), the filling angle decreases (increases) from $\pi - \theta_1$ until the meniscus collapses and the particle is detached from (enclosed by) the interface. In this perspective, $D$ can be regarded as a function of $\psi$. From Eq.~(\ref{eq:interface_height_container}), we get
\begin{align} \label{eq:D_derivative}
 \frac{\mathrm{d} D}{\mathrm{d} \psi} = \frac{\mathrm{d} h}{\mathrm{d} \psi} + \frac{\mathrm{d} \Delta h}{\mathrm{d} \psi} - R \sin \psi~.
\end{align} 
Since $V$ is fixed, $\mathrm{d} V = 0$ and Eq.~(\ref{eq:vol_bath}) yield
\begin{align} \label{eq:derivative_constraint_2}
\pi R^2 \sin^2\psi \frac{\mathrm{d} \Delta h}{\mathrm{d} \psi} + \pi L^2\frac{\mathrm{d} h}{\mathrm{d} \psi} - \pi R^3 \sin^3 \psi = 0~.
\end{align} 
Combining Eqs.~(\ref{eq:D_derivative}) and (\ref{eq:derivative_constraint_2}), we obtain
\begin{align} \label{eq:D_derivative_2}
 \frac{\mathrm{d} D}{\mathrm{d} \psi} =-L^{-2} \left( R^2 \sin^2\psi - L^2\right)
\left(\frac{\mathrm{d} \Delta h}{\mathrm{d} \psi}  - R \sin \psi \right)~.
\end{align} 
As a result, the condition $\frac{\mathrm{d} D}{\mathrm{d} \psi}=0$ when $V$ is fixed is also equivalent to Eq.~(\ref{eq:meniscus_height_derivative}).

The meniscus profile, given by $r(z_0)$, and the meniscus height, $\Delta h$, only depend on $\theta_1$, $R$, $L$, and $\psi$. Furthermore, $\Delta h$ can be written as $R\times f(\theta_1, \psi, L/R)$ as shown in Eq.~(\ref{eq:height_elliptic}). Therefore, the solutions to Eq.~(\ref{eq:meniscus_height_derivative}), $\psi_{\text{min}}$ and $\psi_{\text{max}}$, only depend on $\theta_1$ and $L/R$, not on $D$ and $V$. As a matter of fact, $\psi_{\text{min}}$ and $\psi_{\text{max}}$ are the filling angles at which the meniscus ruptures or collapses, respectively \cite{Anachkov2016}. When a particle is pulled or pushed at a liquid-vapor interface with different $h_0$ (i.e, different $V$ and $D$) but the same $\theta_1$, $R$, and $L$, exactly the same force-displacement curve is expected, as well as the same $\psi_{\text{min}}$ and $\psi_{\text{max}}$. The condition $\frac{\mathrm{d} D}{\mathrm{d} \psi}=0$ was also used previously by Anachkov \textit{et al.} to find the critical angle at which a meniscus breaks \cite{Anachkov2016}.

\begin{figure}[ht]
  \includegraphics[width = 0.35 \textwidth]{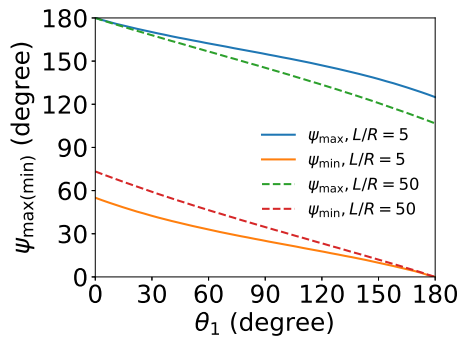}
  \caption{Critical filling angles $\psi_{\text{min}}$ (top two lines) and $\psi_{\text{max}}$ (bottom two lines) vs. $\theta_1$ for $L/R=5$ (solid lines) and 50 (dashed lines).}
  \label{fg:critical_psi}
\end{figure}

The results of the critical filling angles $\psi_{\text{min}}$ and $\psi_{\text{max}}$ as functions of $\theta_1$, determined using Eq.~(\ref{eq:meniscus_height_derivative}), are shown in Fig.~\ref{fg:critical_psi} for $L/R=5$ and 50, respectively. Some interesting behaviors are observed. When $\theta_1 \rightarrow 0$, $\psi_{\text{max}} \rightarrow \pi$ independent of $L/R$. When $\theta_1 \rightarrow \pi$, $\psi_{\text{min}} \rightarrow 0$. Further analysis shows that $\psi_{\text{min}}$ and $\psi_{\text{max}}$ as functions of $\theta_1$ satisfy the following relationship,
\begin{align} \label{eq:psi_min_max}
\psi_{\text{min}}(\theta_1) + \psi_{\text{max}}(\pi-\theta_1) = \pi~.
\end{align} 
This identity originates from the invariance of the Young-Laplace equation [Eq.~(\ref{eq:Young-Laplace})] under the transformation $z_0\rightarrow -z_0$, $\theta_1 \rightarrow \pi - \theta_1$, and $\Delta p\rightarrow-\Delta p$ \citep{Concus1968}.

\begin{figure*}[hbt]
  \includegraphics[width = 1.0 \textwidth]{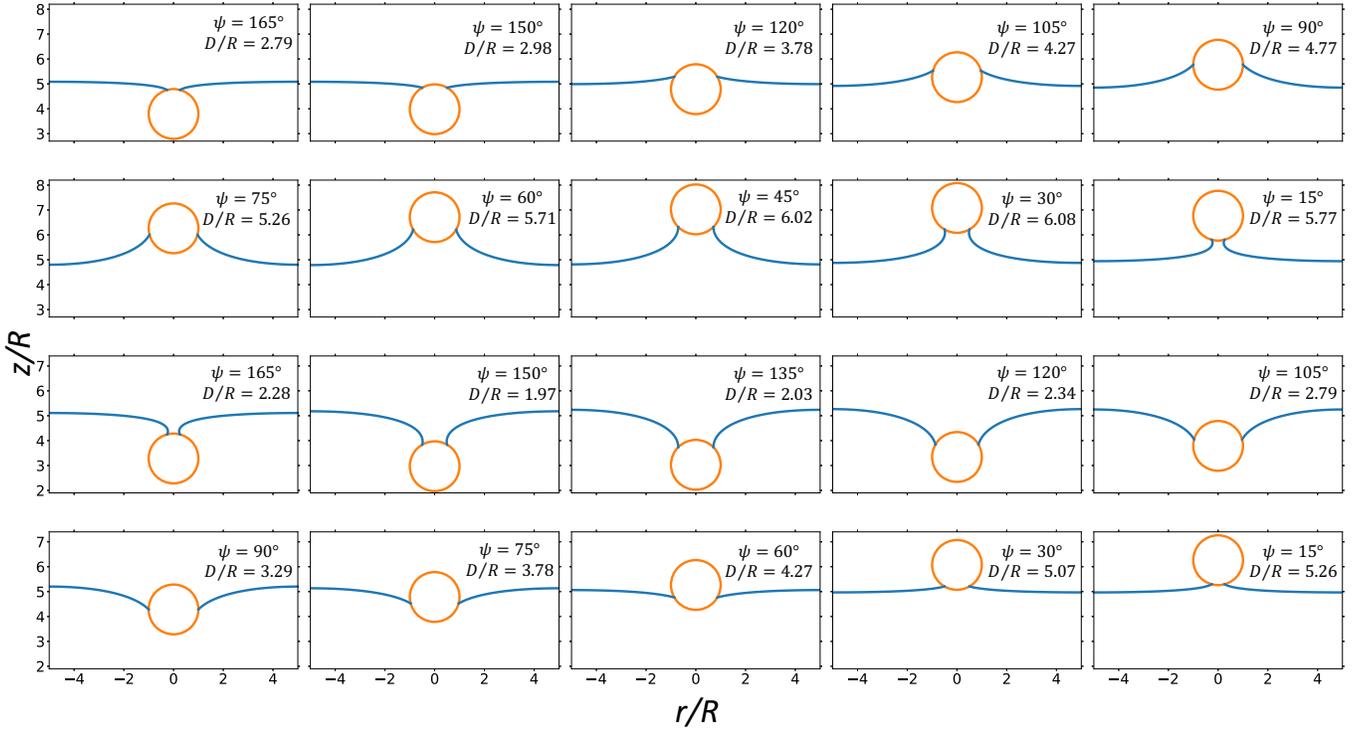}
  \caption{Meniscus profiles predicted by the theory in Sec.~\ref{sec:2b} for $L/R = 5$ and $V/R^3=125\pi$. The first and second rows are for a solvophilic sphere with $\theta_1 = 45^{\circ}$ and $D_0/R=3.34$. The third and fourth rows are for a solvophobic sphere with $\theta_1 = 135^{\circ}$ and $D_0/R=4.71$. The values of $\psi$ and $D$ are indicated in each plot.}
  \label{fg:meniscus_profiles}
\end{figure*}

\subsection{Meniscus profiles for $L \ll \kappa^{-1}$} \label{ss:meniscus_profiles}

With the procedure described in Sec.~\ref{ss:procedure_filling_angle} to determine the filling angle $\psi$, we can theoretically predict the meniscus profile for any given set of $R$, $L$, $\theta_1$, $h_0$, and $D$ (or $\Delta z$) in the limit of $L \ll \kappa^{-1}$. Some examples are shown in Fig.~\ref{fg:meniscus_profiles} for a solvophilic and a solvophobic sphere, respectively. The results indicate that the theory and the procedure presented here can be used to determine the meniscus profile accurately and efficiently for a wide range of parameters and configurations. Below we directly compare the theoretical predictions of the meniscus profile to those obtained from MD simulations and discuss the force-displacement curves in detail.

\begin{figure*}[hbt]
  \includegraphics[width = 0.8 \textwidth]{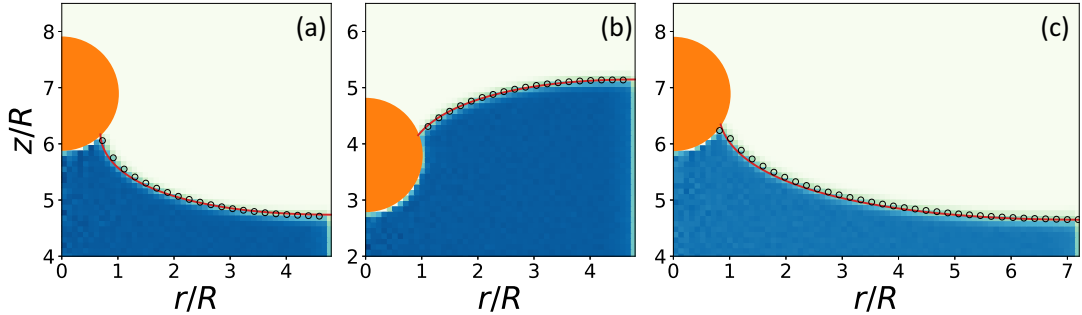}
  \caption{Comparison of the meniscus profile between the theory in Sec.~\ref{sec:2b} and MD simulations for (a) $\theta_1 = 48.5^\circ$, $D = 61\sigma$, $L_\text{n} = 50 \sigma$, and $h_0 = 51.4\sigma$; (b) $\theta_1 = 120.5^\circ$, $D = 29\sigma$, $L_\text{n} = 50 \sigma$, and $h_0 = 51\sigma$; (c) $\theta_1 = 48.5^\circ$, $D = 61\sigma$, $L_\text{n} = 75 \sigma$, and $h_0 = 51.1\sigma$. The angular averaged density of the liquid is represented by a color-scale plot. The black circles indicate the location of the liquid-vapor interface from simulations. The red line indicates the theoretical prediction of the meniscus profile.}
  \label{fg:meniscus_theo_sim}
\end{figure*}

To make a fair comparison between the simulated results and macroscopic theory, care must be taken in defining the radii of the particle and the cylindrical container in the simulations. Repulsive hard-cores of LJ potentials and their integrated forms lead to an excluded zone on any solid surface in which no liquid resides and make the effective radii larger than the nominal radii set in the simulations. We found that the effective radii are $R=10.35 \sigma$ for the particle with $R_\text{n} = 10\sigma$ and $L=49.7 \sigma$ ($74.7\sigma$) for the cylindrical container with $L_\text{n} = 50 \sigma$ ($75 \sigma$). These effective radii are used in the theoretical analyses of the systems simulated with MD.

In simulations the liquid-vapor interface can be located directly from the density distribution of the liquid. Statistical fluctuations of the interface can be reduced numerically by noting the axisymmetry of the systems simulated. Using color-scale plots, Fig.~\ref{fg:meniscus_theo_sim} shows the angle-averaged density profile $\rho(r, z)$ of the liquid as a function of height $z$ and radial distance $r$ from the central axis of the container. The density profiles were averaged over 51 snapshots from MD simulations. We computed the location of the interface at a given $r$ by fitting $\rho(r, z)$ to Eq.~(\ref{eq:interface}). The results are shown in Fig.~\ref{fg:meniscus_theo_sim} as circles. The red lines are the predictions based on Eqs.~(\ref{eq:xt}) and (\ref{eq:yt_elliptic}). In all cases with different $\theta_1$ and $L_\text{n}$, an excellent agreement is found between the simulation and theory, no matter the meniscus is rising [Figs.~\ref{fg:meniscus_theo_sim}(a) and (c)] or depressed [Fig.~\ref{fg:meniscus_theo_sim}(b)]. The good agreement indicates that the particle size is large enough such that possible effects associated with line tensions are negligible and the macroscopic theory of capillarity is applicable for menisci at nanometer scales.

\subsection{Force-displacement curves for $L \ll \kappa^{-1}$}

\begin{figure}[htb]
  \includegraphics[width = 0.48 \textwidth]{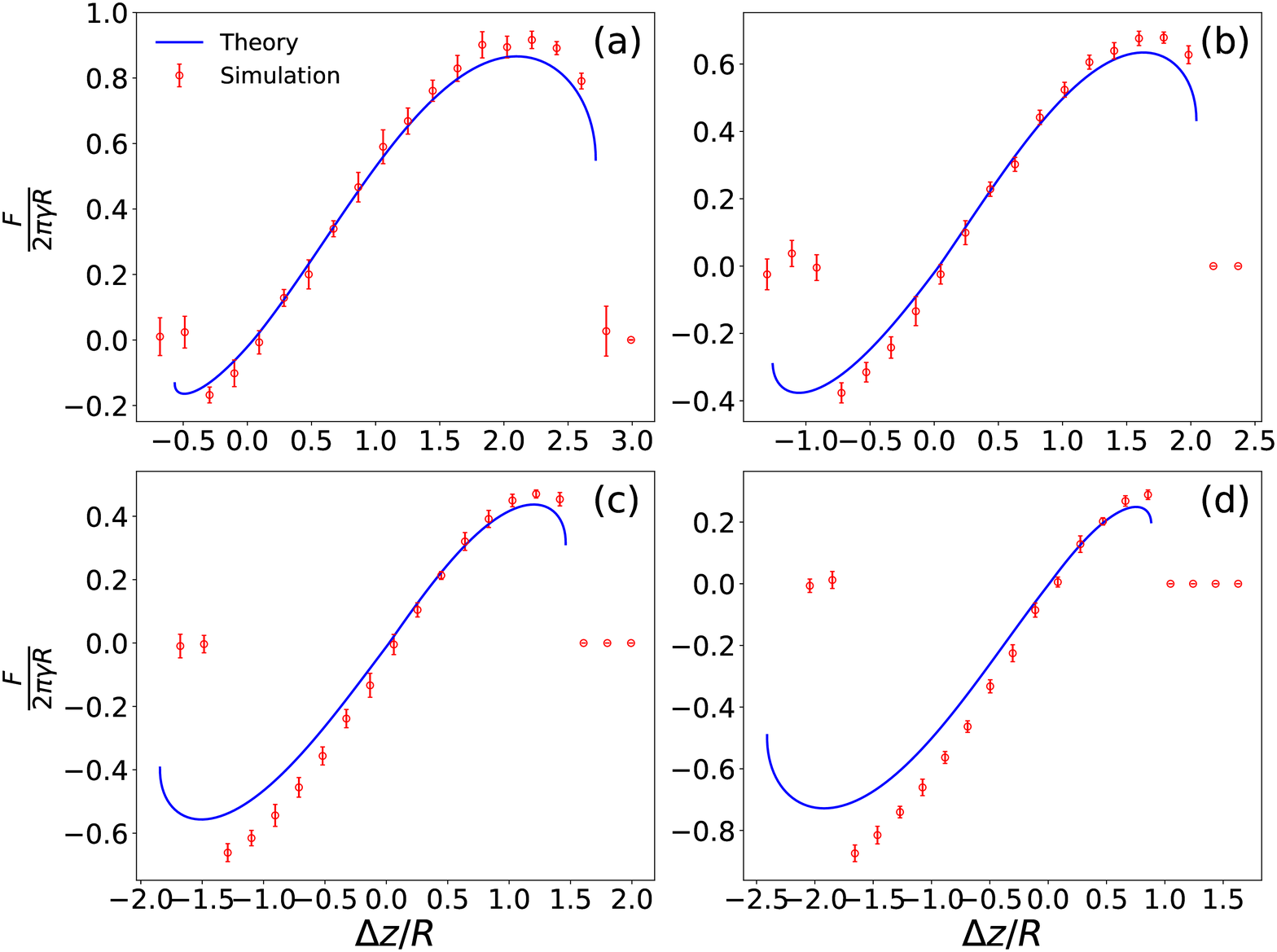}
  \caption{Capillary force ($F$) vs. vertical displacement of the particle ($\Delta z$) from its equilibrium location at a liquid-vapor interface for $L_\text{n} = 50\sigma$ and various values of $\theta_1$: (a) $48.5^\circ$, (b) $76.5^\circ$, (c) $98.2^\circ$, and (d) $120.5^\circ$. The red circles are results from MD simulations and the blues lines are the corresponding theoretical predictions.}
  \label{fg:force_distance_L50}
\end{figure}

Our main goal in this paper is to understand the effective potential confining a particle to its equilibrium location at a liquid-vapor interface. In Fig.~\ref{fg:force_distance_L50}, the force-displacement curves are shown for the systems with $R_\text{n}=10\sigma$ ($R=10.35\sigma$), $L_\text{n} = 50\sigma$ ($L=49.7\sigma$), and various values of $\theta_1$. The symbols represent the capillary force computed in MD simulations using the pulling protocol described in Sec.~\ref{sec:simu}. To quantify the uncertainty of MD calculations, we partitioned the total simulation time ($5000\tau$) during which the force was computed into 10 blocks. An average force was computed for each block. Then the average over all 10 blocks was taken as the final mean force and the standard deviation of the 10 block averages was plotted as error bars in Fig.~\ref{fg:force_distance_L50}.

The solid lines in Fig.~\ref{fg:force_distance_L50} are the predictions of the macroscopic theory of capillarity described in Sec.~\ref{sec:2b}. Since the simulations are in the limit of $R < L \ll \kappa^{-1}$, Eq.~(\ref{eq:force_H}) is used for the capillary force in the theory. In all cases, the theoretical predictions agree reasonably with the MD results, especially in the region of $0 \lesssim \Delta z \lesssim R$. However, the theoretical values tend to be systematically lower than those computed in simulation in terms of magnitude. Clear deviation is observed when the capillary force is close to its extremal values before the meniscus breaks up or the particle submerges into the liquid bath. In particular, the theory seems to work well for solvophilic particles with $\theta_1 < \pi/2$ [Figs.~\ref{fg:force_distance_L50}(a) and (b)]but less so for solvophobic particles with $\theta_1 > \pi/2$ [Figs.~\ref{fg:force_distance_L50}(c) and (d)]. For the latter systems the theoretical predictions of the capillary force are significantly lower than the MD results for the $F < 0$ branch with regard to magnitude, up to about $30\%$ right before particle immersion. At this point there is no physical explanation of the observed discrepancy between the theory and simulations with regard to capillary force, though the two agree very well when meniscus profiles are concerned. One possibility might be when a solvophobic particle is close to its submerging or detaching point, the meniscus strongly bends and the interfacial tension of such a bent interface starts to deviate from the value computed for a flat interface without any curvature \cite{Kashchiev2003}.

\begin{figure}[htb]
  \includegraphics[width = 0.375 \textwidth]{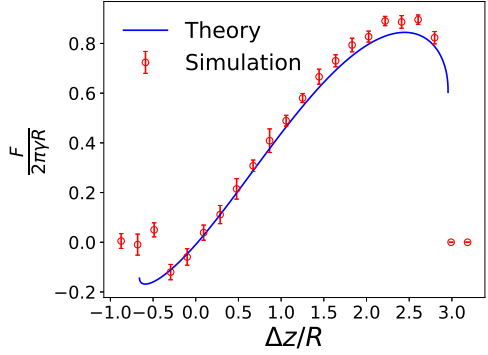}
  \caption{Capillary force ($F$) vs. vertical displacement of the particle ($\Delta z$) from its equilibrium location at a liquid-vapor interface for $L_\text{n} = 75\sigma$ and $\theta_1 = 48.5^\circ$. The red circles are results from MD simulations and the blues line is the corresponding theoretical prediction.}
  \label{fg:force_distance_L75_Ans100}
\end{figure}

Figure ~\ref{fg:force_distance_L75_Ans100} shows the force-displacement curve for a system with $R_\text{n}=10\sigma$ ($R=10.35\sigma$), $L_\text{n} = 75\sigma$ ($L=74.7\sigma$) and $\theta_1 = 48.5^{\circ}$. In this case, the macroscopic theory fits the $F < 0$ branch (i.e., an upward pushing force) well. However, for the region where $F>0$ and the capillary force is pulling the particle downward into the liquid bath, the theoretical prediction is again lower than the simulation results. The largest deviation occurs when the capillary force is near its maximum value at $\Delta z \simeq 2.3 R$. The corresponding rising meniscus breaks up when the particle is pulled upward further.

\begin{figure}[ht]
  \includegraphics[width = 0.48 \textwidth]{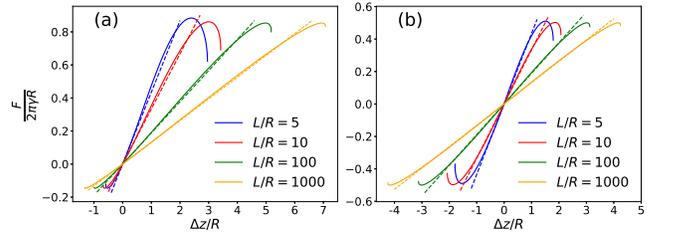}
  \caption{Capillary force ($F$) vs. vertical displacement of the particle ($\Delta z$) from its equilibrium location at a liquid-vapor interface: theory (solid lines) vs. the Joanny-de Gennes' Hookean form in Eq.~(\ref{eq:deGennes-Hooke}) (dashed lines) for (a) $\theta_1=45^\circ$ and (b) $\theta_1=90^\circ$ with $L/R = 5$ (blue), 10 (red), 100 (green), and 1000 (yellow) from most to lest tilted in each plot, respectively.}
  \label{fg:deGennes}
\end{figure}

When the ratio $l\equiv L/R$ gets larger, the dimensionless mean curvature of the interface becomes smaller roughly as $l^{-2}$ [see Eq.~(\ref{eq:H})]. Eventually the meniscus profile reduces to a catenary curve with zero mean curvature \cite{DeGennes2004,Tang2018arXiv}. The meniscus height for a catenoid is very similar to the expression in Eq.~(\ref{eq:height_logL}) but without the -1/2 term in the curly brackets. In the limit of $L \gg R$, the force-displacement curve reduces to the Joanny-de Gennes' law in Eq.~(\ref{eq:deGennes-Hooke}) with an effective spring constant $k_s = 2\pi\gamma/\ln(2L/R)$ \cite{Joanny1984}. In Fig.~\ref{fg:deGennes} this linear force-displacement relationship is compared to the theoretical solutions using the elliptic integrals for $\theta_1 = 45^\circ$. Even for $L/R=5$, the theoretical results fit reasonably to the Joanny-de Gennes' law with a linear behavior apparent for $|\Delta z/R | \ll1$, though deviations can be seen at lager displacements or when the particle is close to detaching from or submerging into the liquid. The agreement is improved and the linear region of the force-displacement curve is widened when $L/R$ becomes larger. For $L/R=1000$, the linear force-displacement curve from the Joanny-de Gennes' law overlaps with the theoretical solution based on the elliptic integrals for a wide range of $\Delta z$, except very close to the extrema at which the capillary force bends and deviates from the linear dependence on $\Delta z$.

\begin{figure}[hbt]
  \includegraphics[width = 0.48 \textwidth]{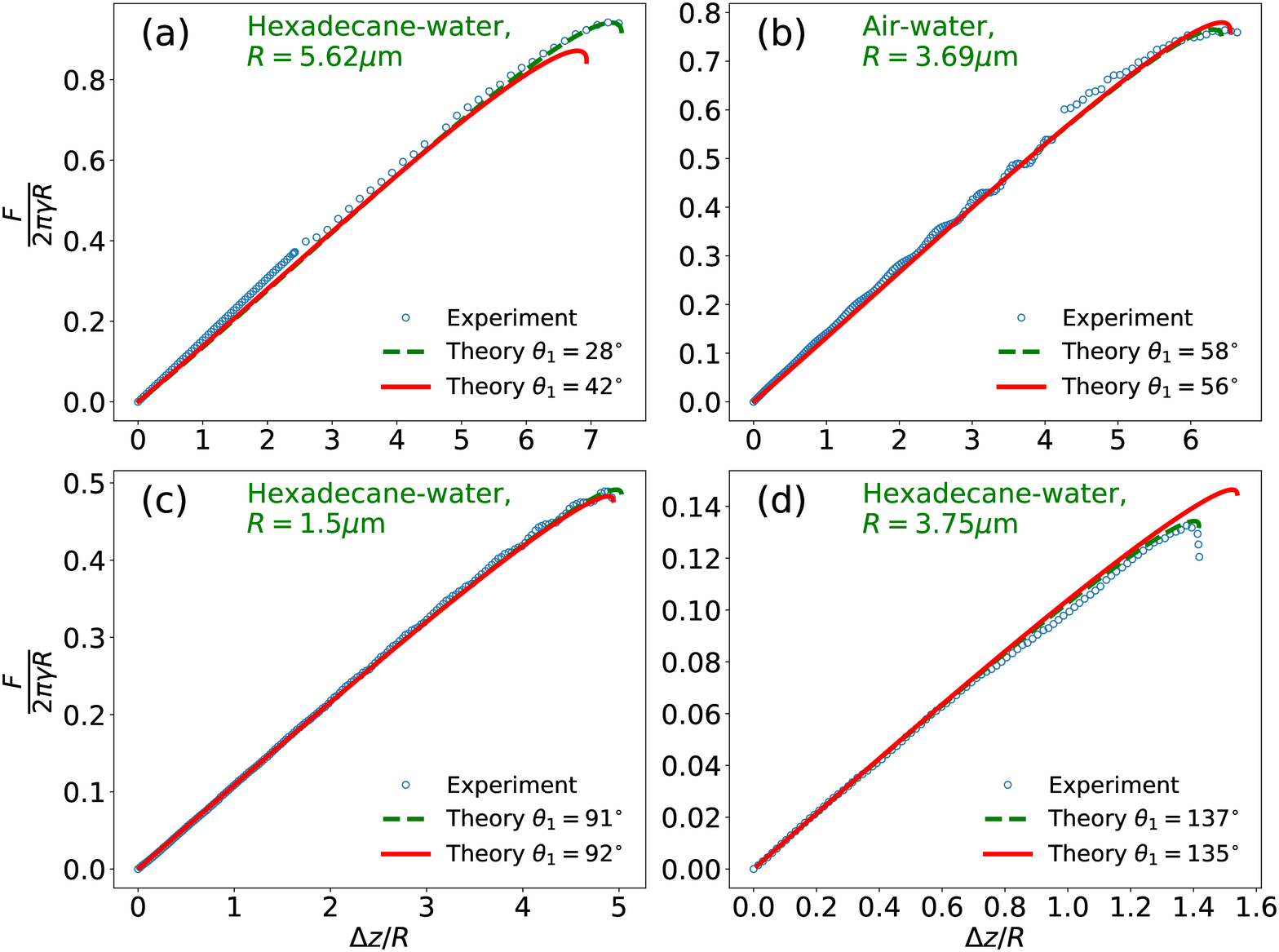}
  \caption{Capillary force ($F$) vs. vertical displacement of the particle ($\Delta z$) from its equilibrium location at a liquid-vapor interface: theory (lines) vs. experimental data (circles, by courtesy of Dr. Anachkov \cite{Anachkov2016}). Panels (a), (c), and (d) are for particles with various radii at a hexadecane-water interface; panel (b) is for a water-air interface. The solid and dashed lines are from the theory in Sec.~\ref{sec:2b} with different values of $\theta_1$ and (a) $L=4$ mm, (b) $L=4$ mm, (c) $L=9$ mm, and (d) $L=17$ mm, respectively.}
  \label{fg:force_exp}
\end{figure}

The approximate linear dependence of the capillary force ($F$) on the particle displacement ($\Delta z$) has been confirmed experimentally. For example, Fig.~\ref{fg:force_exp} includes some experimental data from the group of Dr. Anachkov for microparticles at water-air and water-oil interfaces, where a linear relationship between $F$ and $\Delta z$ is apparent unless the meniscus is close to breaking up \cite{Anachkov2016}. In the experimental setup to measure $F$, the liquid (water) was filled in a cone-shaped container in the middle of a much larger vessel. Water and air or oil met at the opening of the cone, the radius of which was about 1 mm. Therefore, the lateral span of the water-air (oil) interface was about 1 mm.

The interfacial tension of a water-hexadecane interface is $52.5 \pm 0.5$ mN/m and of a water-air interface is 72 mN/m at 25 $^{\circ}$C. The density of hexadecane is 770 kg/m$^{3}$ and that of water is $1000$ kg/m$^{3}$. As a result, the capillary length ($\kappa^{-1}$) is about 4.82 mm and 2.7 mm for a water-hexadecane and a water-air interface, respectively. In the experiment, the radius of the particle was varied from 1.5 $\mu$m to $5.62$ $\mu$m, much smaller than $\kappa^{-1}$. Our previous work showed that the theory presented in Sec.~\ref{sec:2b} and the solutions of the Young-Laplace equation based on the elliptic integrals apply up to about $L \sim 0.4\kappa^{-1}$, which is about 1.9 mm for the water-hexadecane and 1.1 mm for the water-air system, respectively. Therefore, the interface involved in the experiments can be assumed to have a constant mean curvature and gravity can be ignored. However, the contact angle of the interface at the edge of the cone, $\theta_2$, is unknown. The theory presented in this paper is based on $\theta_2=\pi/2$. In order to use the theory in Sec.~\ref{sec:2b} to fit the experimental data, we will treat $L$ as a fitting parameter in the theory. This treatment is based on the assumption that a meniscus emerging in the experiments with $L = 1$ mm and $\theta_2 \neq \pi/2$ is only a portion of a meniscus with a different $L$ but $\theta_2 = \pi/2$.

The lines in Fig.~\ref{fg:force_exp} are the theoretical fits using Eqs.~(\ref{eq:height_elliptic}), (\ref{eq:force_H}), and (\ref{eq:delta_z}) with $L$ as a fitting parameter. An excellent agreement is found between the theory and experimental results for all the cases. The two lines are for two contact angles that were reported in Ref.~\cite{Anachkov2016} for each case using different measurement techniques. It should be noted that almost identical fits can be obtained using Eqs.~(\ref{eq:height_logL}), (\ref{eq:force_approx}), and (\ref{eq:delta_z_delta_h}). This fact is not a coincidence and can be understood as follows. In the experiments $L \gg R$ and the mean curvature of the meniscus is thus close to 0. In this limit, the meniscus is essentially a catenoid, to which Eqs.~(\ref{eq:height_logL}), (\ref{eq:force_approx}), and (\ref{eq:delta_z_delta_h}) apply.

In Ref.~\cite{Anachkov2016}, the same experimental data included in Fig.~\ref{fg:force_exp} were fit using the Derjaguin-James formula [Eq.~(\ref{eq:height_DJ})] for the meniscus rise. Some deviations were noted. The Derjaguin-James formula was derived for a meniscus with an unbound lateral span. Our previous work showed that it applies when $L \gtrsim 4\kappa^{-1}$ for $\theta_2=\pi/2$ \cite{Tang2018arXiv}. These conditions were not met in the experiments, which may explain the observed difference between the fits using the Derjaguin-James formula and the experimental data \cite{Anachkov2016}.

\subsection{Effects of gravity for $L \gtrsim \kappa^{-1}$}

In Sec.~\ref{sec:2b}, it is found that a meniscus on the outside of a particle with $R \ll L \ll \kappa^{-1}$ is a surface of revolution with a constant mean curvature. In this case, the meniscus height ($\Delta h$) grows with $L$ logarithmically, as shown in Eq.~(\ref{eq:height_logL}). When $L$ becomes comparable to or larger than $\kappa^{-1}$, gravity comes into effect and the logarithmic growth ceases with $\Delta h$ saturating to a value predicted by the Derjaguin-James formula [Eq.~(\ref{eq:height_DJ})]. The logarithmic dependence of $\Delta h$ on $L$ is the origin of the Joanny-de Gennes' law in Eq.~(\ref{eq:deGennes-Hooke}) for $R \ll L \ll \kappa^{-1}$, which states that the effective spring constant associated with a liquid-vapor interface for a particle straddling the interface can be written as $k_s \simeq 2 \pi \gamma/\ln(2L/R)$. According to this expression, $k_s$ gradually decreases as $L$ increases. However, as $\Delta h$ eventually saturates and becomes $L$-independent when $L \gg \kappa^{-1}$, the spring constant is expected to saturate in the same limit.

From our previous work on the wetting of a cylinder vertically penetrating a liquid-vapor interface, we know that Eq.~(\ref{eq:height_elliptic}) is accurate for the meniscus height for $L$ up to $0.4\kappa^{-1}$ with the parameter $l\equiv L/R$. When $L > 1.85\kappa^{-1}$, Eq.~(\ref{eq:height_elliptic}) can still be used for $\Delta h$ with the parameter $l$ replaced by $1.85\kappa^{-1}/R$. This finding leads to a re-definition of $l$ in Eq.~(\ref{eq:l_parameter}), which can be combined with Eq.~(\ref{eq:height_elliptic}) and an error correcting function [Eq.~(\ref{eq:fit_kink_function})] to yield an approximate formula of $\Delta h$. The resulting formula is shown in Eq.~(\ref{eq:height_final}) and works for an arbitrary $L$ including the crossover zone $0.4\kappa^{-1} \lesssim L \lesssim 4\kappa^{-1}$.

\begin{figure}[ht]
  \includegraphics[width = 0.4 \textwidth]{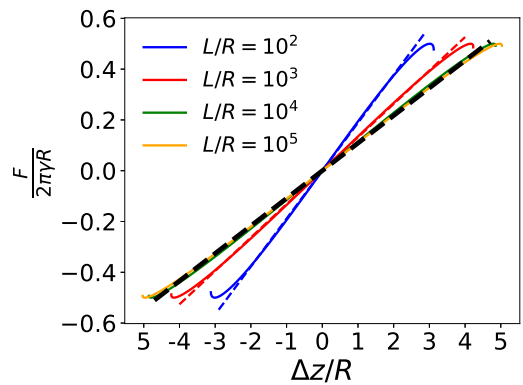}
  \caption{Capillary force ($F$) vs. vertical displacement of the particle ($\Delta z$) from its equilibrium location at a water-air interface: theory (solid lines) vs. the Hookean form with the effective spring constant in Eq.~(\ref{eq:spring_constant}) (dashed lines) for $\theta_1=90^\circ$, $R=1~\mu$m, and $L/R = 10^2$ (blue), $10^3$ (red), $10^4$ (green), and $10^5$ (yellow) from most to lest tilted, respectively. In particular, the black dashed line corresponds to $k_s = 2 \pi \gamma/\ln (3.7\kappa^{-1}/R)$, which is the saturated value of $k_s$ for large $L$.}
  \label{fg:hookean_g}
\end{figure}

Similar to $\Delta h$, the spring constant $k_s$ also exhibits crossover and saturation when the lateral span of the liquid-vapor interface is varied. As result, we can generalize the expression of $k_s$ as
\begin{align} \label{eq:spring_constant}
k_s \simeq \frac{2 \pi \gamma}{\ln(2l)}~,
\end{align}
with the parameter $l$ given in Eq.~(\ref{eq:l_parameter}) instead of being always $L/R$. This expression of $k_s$ is expected to work for an arbitrary value of the ratio $L/R$. In Fig.~\ref{fg:hookean_g} we plot the force-displacement curves for a particle at a water-air interface computed using Eqs.~(\ref{eq:height_final}), (\ref{eq:delta_z_largeL}), and (\ref{eq:force_approx_2}) for $\theta_1=\pi/2$, $R=1~\mu$m and various values of $L$ ranging from 100 $\mu$m to 100 mm. The capillary length is $\kappa^{-1} = 2.7$ mm. The linear region of each curve has a slope that agrees well with the effective spring constant from Eq.~(\ref{eq:spring_constant}), i.e.,$k_s = 2 \pi \gamma/\ln (2l)$ with $l = L/R$ for $L \le 1.85\kappa^{-1}$ while $l = 1.85\kappa^{-1}/R$ for $L > 1.85\kappa^{-1}$. It is clear that when $L\rightarrow \infty$, the spring constant $k_s$ saturates to $2 \pi \gamma/\ln (3.7\kappa^{-1}/R)$.

\section{Conclusions} \label{sec:conclusions}

In this paper, we present a comprehensive theory of the meniscus on the outside of a small particle (i.e., $R\ll \kappa^{-1}$ with $\kappa^{-1}$ being the capillary length of the interface involved) at a liquid-vapor interface confined in a cylindrical container with radius $L$ ($>R$). By placing the particle along the central axis of the container, we computed the capillary force on the particle when it was displaced out of its equilibrium height relative to the interface. With the contact angle of the liquid on the container surface being fixed at $\pi/2$, the setup allowed us to systematically study the crossover from a meniscus with a constant Laplace pressure to an unbound one governed by gravity, when $L$ is increased from $L\ll \kappa^{-1}$ to $L\gg \kappa^{-1}$. In the limit of $R<L\ll \kappa^{-1}$, an analytic solution based on the elliptic integrals was found for the Young-Laplace equation, resulting in a meniscus of a constant mean curvature and with a height that grows logarithmically with $L$. In the limit of $R \ll \kappa^{-1} \ll L$, the meniscus height saturates and is given by the Derjaguin-James formula. In the crossover region, which is roughly $0.4 \kappa^{-1} \lesssim L \lesssim 4\kappa^{-1}$, an approximate formula is proposed for the meniscus height based on our previous work on the wetting of a cylinder.

MD simulations show that the meniscus shape at nanometer scales matches well the prediction of the macroscopic theory of capillarity based on the Young-Laplace equation. The capillary force is reasonably predicted by the theory as well, especially the branch where the force is attractive and pulling the particle toward the liquid phase. However, for the repulsive branch where the force is pushing the particle out of the liquid, the theory always predicts a force with a magnitude smaller than that computed in MD simulations. The origin of this discrepancy is not understood at present.

The simulation and theoretical results show that the capillary force on a small particle at a liquid-vapor interface can be reasonably approximated as a linear function of the displacement of the particle out of its equilibrium location, especially for small deviations. This approximation holds from nanometer to macroscopic scales and the associated effective spring constant can be written as $k_s = 2 \pi \gamma / \ln (2l)$. For $L \le 1.85\kappa^{-1}$, the parameter $l = L/R$, indicating that $k_s$ decreases as the reciprocal of $\ln L$ as $L$ increases. For $L > 1.85\kappa^{-1}$, $k_s$ saturates to $2 \pi \gamma / \ln (3.7\kappa^{-1}/R)$.

Our result on $k_s$ differs from the result of Pieranski \cite{Pieranski1980}, who predicted $k_s = 2 \pi \gamma$, by the factor $\ln^{-1} (2l)$ associated with the lateral span of the meniscus. In Pieranski's analyses, the liquid-vapor interface was always flat regardless of the location of the particle's center relative to the interface. In other words, the deformation of the meniscus was not considered when the particle was displaced out of its equilibrium location. The analyses was based on free energies but not self-consistent as the force exerted on the particle by the liquid-vapor interface was always 0. In this paper, we fully account for the evolution of the meniscus on the outside of a particle moving across a liquid-vapor interface. Our results thus provide a rigorous theoretical foundation of the moving interface method in which a liquid-vapor interface is modeled as a harmonic potential with regard to its confining effect on small particles at the interface or in the liquid phase. A physical interpretation is found for the associated spring constant in terms of the surface tension of the interface ($\gamma$),  the particle size ($R$), the lateral span of the interface ($L$), and possibly the capillary length ($\kappa^{-1}$) of the interface when $L$ is large. We expect this approach of modeling a liquid-vapor interface as a confining potential for particles will find wide applications in simulating evaporation of particle suspensions, interfacial adsorption and assembly of particles, and many other processes involving particles at interfaces.

We derive $k_s$ using a single particle at a liquid-vapor interface. When multiple particles are adsorbed at an interface, capillary interactions mediated by menisci can occur. In this case, it is nontrivial to model the interface as a confining potential for each particle. However, if we interpret $L$ as the average interfacial distance between the particles, similar to the treatment of Joanny and de Gennes of a contact line pinned by multiple defects \cite{Joanny1984}, then a harmonic potential with $k_s = 2 \pi \gamma / \ln (2l)$ for each particle at the interface will at least partially capture the capillary interactions.

\begin{acknowledgments}
Acknowledgment is made to the Donors of the American Chemical Society Petroleum Research Fund (PRF \#56103-DNI6), for support of this research. S.C. thanks the partial support from a 4-VA Collaborative Research Grant (``Short-range interactions of interfacial bubbles''). The authors thank Dr. Anachkov for sharing their experimental data.
\end{acknowledgments}

%\bibliography{tetramer}

%merlin.mbs apsrev4-1.bst 2010-07-25 4.21a (PWD, AO, DPC) hacked
%Control: key (0)
%Control: author (0) dotless jnrlst
%Control: editor formatted (1) identically to author
%Control: production of article title (0) allowed
%Control: page (1) range
%Control: year (0) verbatim
%Control: production of eprint (0) enabled
%

\end{document}